\documentclass[12pt]{iopart}

\usepackage{graphicx}
\usepackage{xcolor}
\begin{document}

\title[Understanding Galaxy Rotation Curves with Verlinde's Emergent Gravity]{Understanding Galaxy Rotation Curves with Verlinde's Emergent Gravity}

\author{Youngsub Yoon$^1$, Jong-Chul Park$^1$, Ho Seong Hwang$^{2,3}$}

\address{$^1$Department of Physics and Institute of Quantum Systems (IQS),\\ Chungnam National University, Daejeon 34134, Republic of Korea\\
$^2$Astronomy Program, Department of Physics and Astronomy,\\ Seoul National University, Seoul 08826, Republic of Korea\\
$^3$SNU Astronomy Research Center,\\ Seoul National University, Seoul 08826, Republic of Korea}
\ead{youngsuby@gmail.com,jcpark@cnu.ac.kr,hhwang@astro.snu.ac.kr}
\vspace{10pt}
\begin{indented}
\item[]August 2022
\end{indented}

\begin{abstract}
We present the results from the analysis of galaxy rotation curves with Verlinde's emergent gravity.
We use the data in the SPARC ({\it Spitzer} Photometry and Accurate Rotation Curves) database,  which contains a sample of 175 nearby disk galaxies with 3.6 $\mu$m surface photometry and rotation curves. 
We compute the gravitational acceleration at different galactocentric radii expected from the baryon distribution of the galaxies with the emergent gravity, and compare it with the observed gravitational acceleration derived from galactic rotation curves. 
The predicted and observed accelerations agree well with a mean offset $\mu{\rm [log(g_{obs})-log(g_{Ver})]}=-0.060\pm0.004$ and a scatter $\sigma{\rm [log(g_{obs})-log(g_{Ver})]}=0.137\pm0.004$ by assuming a de Sitter universe.
These offset and scatter become smaller when we assume a more realistic universe, quasi de Sitter universe, as $\mu=-0.027\pm0.003$ and $\sigma=0.129\pm0.003$.
Our results suggest that Verlinde's emergent gravity could be a good solution to the missing mass problem without introducing dark matter.
\end{abstract}

%
%
%
%
%

\section{Introduction}\label{introduction}

The missing mass problem, first reported in the 1930s~\cite{Zwicky:1933gu}, is now very well established~\cite{Trippe:2014hja}.
There have been two major approaches to solve this problem.
The first one is {\it dark matter}, which asserts that there is invisible extra matter.
The second one is to assume that the Newton-Einstein gravity theory is incorrect.

The first approach, most notably, $\Lambda$ Cold Dark Matter ($\Lambda$CDM) model achieved a great success, despite some discrepancies~\cite{Bullock:2017xww} including the Hubble tension~\cite{DiValentino:2021izs}.
The second approach has been less successful, but its most notable theory,  Modified Newtonian Dynamics (MOND) could explain a scaling relation for galaxies: e.g. Tully-Fisher relation, an empirical correlation between the intrinsic luminosity (or mass) of a disk galaxy and the emission line width (i.e., rotation velocity)~\cite{Tully:1977fu, Milgrom:1983ca}.
On the other hand, the dark matter theory seems to have some difficulties in explaining the relation quantitatively~\cite{Lelli:2022hvc}.

MOND has been used to successfully fit the galaxy rotation curves with a relatively small scatter~\cite{McGaugh:2016leg}.
Considering relatively large  observational errors for the gravitational acceleration, it is a remarkable result~\cite{Lelli:2016cui}.
Despite this success, MOND is not fully satisfactory.
This is because a free parameter called ``Milgrom's constant'' should be tuned to fit the rotation curves.
Moreover, the function that fits the rotation curves does not seem to have any good physical basis.

In this regard, the emergent gravity that Verlinde proposed looks more promising~\cite{Verlinde:2016toy}.
There is no free parameter whose sole purpose is to fit the galaxy rotation curves.
Actually, there is a new parameter that is used to fit the rotation curves.
However, the parameter comes from the cosmological observations of the expansion of the universe, which is independent of the rotation curve measurements.
Moreover, unlike MOND, the function that fits the rotation curves is naturally derived from simple physical principles. Thus, Verlinde's emergent gravity is advantageous over other modified gravity theories whose Lagrangians are deliberately constructed to reproduce the Tully-Fisher relation.
Despite these advantages, it has been difficult to fit the rotation curves with Verlinde's emergent gravity because of the lack of its generalization to the case of the absence of the spherical symmetry in the mass distribution.
Recently, there was a progress in this direction~\cite{Yoon:2021bse}, which enabled this study. Therefore, in this article, we will examine whether we can explain the galaxy rotation curves with Verlinde's emergent gravity without introducing dark matter.

The organization of this paper is as follows. In Section~\ref{TullyFisherMOND}, we review the Tully-Fisher relation and illuminate how MOND explains it. In Section~\ref{Verlindegravity}, we review the emergent gravity to provide the background on the theory and the equations, which we compare the galaxy rotation curves with.
In Section~\ref{methods}, we explain the method we use for the analysis of the data and analysis results.
In Section~\ref{discussion}, we discuss our results and conclude our paper.

\section{The Tully-Fisher relation and MOND}\label{TullyFisherMOND}

We first review how to explain the Tully-Fisher relation with MOND to better understand why replacing the Newtonian gravity by another theory of gravity appears more promising for explaining the galaxy rotation curves. 
The observed Tully-Fisher relation suggests that the speed of the outermost stars orbiting around the center of the galaxy ($v$) is proportional to the fourth root of the total mass of stars in the galaxy ($M$), and is independent of the distance from the center as long as they are far enough from the center~\cite{Tully:1977fu}.
In other words, we have
\begin{equation}
v \propto M^{1/4}\,.\label{TullyFisher}
\end{equation}
It is difficult to explain this relation with the Newtonian gravity without assuming dark matter (i.e., assuming that only the baryonic (or visible) matter contributes to gravity)~\cite{Lelli:2015wst}.
According to the Newtonian gravity, the rotation velocity is expected to decrease as the star is farther from the center, and the rotation velocity is proportional to $M(r)^{1/2}$ as follows
\begin{equation}
v=\sqrt{\frac{GM(r)}{r}}\,,    
\end{equation}
where $M(r)$ is an enclosed mass at radius $r$.
This relation is different from the Tully-Fisher relation (\ref{TullyFisher}) in a sense that the velocity of outermost stars is proportional to $M^{1/2}$ instead of $M^{1/4}$, and decreases as $r^{-1/2}$ when $r$ is large enough to regard $M(r)$ as constant. 

As an alternative to dark matter, Mordehai Milgrom proposed MOND~\cite{Milgrom:1983ca}, which can explain the Tully-Fisher relation.
In particular, he proposed the following modification to Newton's second law:
\begin{equation}
F=m\frac{a^2}{a_M}\quad {\rm for}\quad a \ll a_M\,,   \label{FMOND}
\end{equation}
where $a_M$ is Milgrom's constant, known to be around $1.2\times 10^{-10}\, \mathrm{m~s^{-2}}$.
Then, we have  
\begin{equation}
m\frac{a^2}{a_M}=\frac{GM(r)m}{r^2},
\end{equation}
which yields
\begin{equation}
a=\frac{\sqrt{GM(r)a_M}}{r}=\frac{v^2}{r}.
\end{equation}
This results in the Tully-Fisher relation, $v=(GMa_M)^{1/4}$.
Actually, the original formulation of the Tully-Fisher relation is given by $v\propto L^{1/4}$ where $L$ is the luminosity of a galaxy.
The luminosity of a galaxy is more or less proportional to the mass of stars in a galaxy, thus we recover Eq.~(\ref{TullyFisher}).
However, stars are not the only constituents of galaxies.
This means that in light of MOND, the Tully-Fisher relation should hold better if the total mass of baryons including stars and gas is plugged in the mass term $M$ of Eq.~(\ref{TullyFisher}) rather than only the stellar mass.
Indeed, this was confirmed in~\cite{McGaugh:2000sr}, which is known as {\it Baryonic Tully-Fisher relation}.

If we denote the observed gravitational acceleration by $g_{\mathrm{obs}}$ and the Newtonian gravitational acceleration from baryons by $g_{\mathrm{bar}}$, Milgrom's proposal can be rewritten as
\begin{equation}
g_{\mathrm{obs}} = \mathcal F(g_{\mathrm{bar}}) =
\left\{\begin{array}{l}
g_{\mathrm{bar}}~\quad\qquad(\mathrm{for}~g_{\mathrm{bar}}\gg a_M)\\
\sqrt{g_{\mathrm{bar}}{a_M}}\quad(\mathrm{for}~g_{\mathrm{bar}}\ll a_M)\,.
\end{array}
\right.
\label{gobsconditions}
\end{equation}
The first condition is necessary because MOND must be reduced to the Newtonian gravity in our daily life, while the second, to satisfy the Tully-Fisher relation.
Note that MOND does not predict the exact form of $\mathcal F(g_{\mathrm{bar}})$ other than these two limits.
Therefore, different arbitrary trials for $\mathcal F(g_{\mathrm{bar}})$ have to be made so that its prediction can be compared with observational data.
In Section \ref{results}, we will present the functional form of $\mathcal F(g_{\mathrm{bar}})$ that is considered for the relation between $g_{\mathrm{bar}}$ and $g_{\mathrm{obs}}$ in Ref.~\cite{McGaugh:2016leg}.

There have been some attempts to understand the Tully-Fisher relation from dark matter by making certain assumptions in galaxy formation scenarios~\cite{Dalcanton:1996de, Mo:1997vb, Mo:2000pu}. 
However, they all obtained $\beta=3$ for $v^\beta \propto M$, which differs from the value from observations, $\beta\simeq4$.
Some observations indeed reported lower values for $\beta$, but they are mainly for dwarf galaxies~\cite{Begum:2008gn, Kirby:2012sg, McCall:2012ya, Brook:2016, Karachentsev:2016cgo}.
This difference can be understood by different gas fractions for dwarf galaxies, and can be removed if we replace the stellar mass with total baryonic mass (i.e., stellar plus gas mass).
However, there are still problems remaining for dark matter, mainly for the scatter and slope of the Tully-Fisher relation as highlighted in Refs.~\cite{Lelli:2015wst, Lelli:2022hvc}.   

On the other hand, despite the success of MOND in explaining the Tully-Fisher relation, there have been some indications that MOND may not be the ultimate answer. For example, the total dynamical mass estimate from Alzain's version of MOND is much smaller than the baryon mass in almost all galaxy clusters~\cite{Seeram:2021xsi}. Otherwise, a variant model of MOND is required with a value of $a_M$ two times smaller, which could be problematic as $a_M$ is already fixed by the Tully-Fisher relation. Also, there is indication that one still needs cold dark matter in the cosmological fluctuations, even in the relativistic generalizations of MOND~\cite{Clifton:2011jh}.

\section{Verlinde's Emergent Gravity}\label{Verlindegravity}

In 2011, Verlinde proposed {\it entropic gravity}~\cite{Verlinde:2010hp}.
He derived Newton's universal gravitation law and Einstein equations assuming that the entanglement entropy in a gravitational system is given by the Bekenstein-Hawking entropy, which follows the area law.
Then, the inverse square law follows from the fact that an area scales as distance squared.
In 2017, the proposal was developed to {\it emergent gravity} assuming an additional volume contribution to the entanglement entropy~\cite{Verlinde:2016toy}.
Therefore, the emergent gravity deviates from the Newtonian gravity at large distances, where the volume contribution is non-negligible compared to the Bekenstein-Hawking entropy.
From this assumption, Verlinde successfully derived the Tully-Fisher relation, including Milgrom's constant. 

The gravitational acceleration in the emergent gravity is given by~\cite{Yoon:2020meh}
\begin{equation}
g_{\mathrm{Ver}}=\sqrt{g_B^2+g_D^2}\label{gVerlinde}\,,
\end{equation}
where $g_B$ is the Newtonian gravity from baryonic matter (i.e., $g_{\mathrm{bar}}$ in Section~\ref{TullyFisherMOND}) and $g_D$ is the gravity from {\it apparent dark matter}.
Apparent dark matter does not exist, but Verlinde named it so because it plays the same role as what dark matter does in the dark matter theory.

Actually, Verlinde suggested $g=g_B+g_D$ in his original paper~\cite{Verlinde:2016toy}.
However, if we consider the fact that the total gravitational energy is the sum of the one due to Newtonian gravity $g_B$ and the one due to $g_D$, we arrive at Eq. (\ref{gVerlinde}), as gravitational energy is proportional to the gravitational field squared. Moreover, $g=g_B+g_D$ is already ruled out by the equivalence principle violation experiments~\cite{Yoon:2021bse}.

The original formulation of Verlinde's gravity is based on the case when there is a spherical symmetry in the matter distribution.
The formulation is extended for the general case~\cite{Yoon:2021bse}, which can be summarized as
\begin{equation}
\Phi_B=-\frac{2g_B^2}{4\pi G \rho_B+\vec n \cdot \nabla g_B}\,,\quad  g_D^2=\frac{a_0}{6}(\vec n \cdot \nabla \Phi_B+2g_B)\,,
\end{equation}
where $\rho_B$ is the baryonic matter density, $\vec n$ is the direction of the Newtonian gravity, and $a_0=c H_0$ for de Sitter universe, which expands exponentially as follows\footnote{Strictly speaking, even when we assume de Sitter universe, this formula is valid only in the case when our universe is flat. However, it is still a good approximation because the curvature of our universe is negligible. Instead of an exponential function, the scale function is $\cosh$ in case of the closed universe, and $\sinh$ in case of the open universe. They are essentially the same as the exponential function for the current curvature of our universe.}:
\begin{equation}
\frac{\dot a}{a}=H_0\,,\qquad  a=e^{H_0 t}\,.
\end{equation}
Here $H_0$ is the Hubble parameter at the current epoch, which we will call the Hubble constant.
In galaxies, $\vec n$ is given by $-\hat r$.
Thus, we obtain
\begin{equation}
\Phi_B=-\frac{2g_B^2}{4\pi G \rho_B-\partial_r g_B}\,,\quad  g_D^2=\frac{a_0}{6}(2g_B-\partial_r \Phi_B)\,. \label{Verlindegalaxy}
\end{equation}

We then can have the Tully-Fisher relation as follows. 
When $\rho_B=0$ and $g_B=GM/r^2$, it is easy to check that the above formulae reduce to
\begin{equation}
\Phi_B=-\frac{GM}{r}\,,~~ g_D^2=\frac{a_0}{6}\left(2g_B-\frac{GM}{r^2}\right)=\frac{a_0}{6} g_B\,.
\end{equation} 
Thus, when $g_B\ll a_0/6$, 
Eq.~(\ref{gVerlinde}) becomes
\begin{equation}
g_{\mathrm{Ver}}\approx \sqrt{\frac{a_0}{6} g_B}\,.
\end{equation}
This is exactly the second condition in Eq.~(\ref{gobsconditions}) with the identification $a_M=a_0/6$.
Plugging in $g_B=GM/r^2$ and $g_{\mathrm{Ver}}=v^2/r$, we obtain
\begin{equation}
\frac{v^2}{r}=\sqrt{\frac{a_0}{6}\frac{GM}{r^2}}
\end{equation}
which reduces to the Tully-Fisher relation.
If we adopt $H_0=70$ km s$^{-1}$ Mpc$^{-1}$, we have
\begin{equation}\label{a6}
a_0/6=cH_0/6=1.13\times 10^{-10}~\mathrm{m~s^{-2}}
\end{equation}
which agrees with Milgrom's constant within 10\%.
This means that Milgrom's constant can be derived from the Hubble constant in the context of Verlinde's emergent gravity.
Note that the factor 1/6 is not arbitrary, but is derived from the theory of emergent gravity.

On the other hand, the $a_0$ value is based on the assumption that our universe is a de Sitter one, which is however not true; we know that the Hubble parameter has not been the same during the history of our universe.
Therefore, the authors of Ref.~\cite{Diez-Tejedor:2016fdn} considered what they called {\it quasi de Sitter universe}.
The only difference is that they considered the quasi de Sitter acceleration scale $a_0=5.41\times 10^{-10}~\mathrm{m~s^{-2}}$ which comes from current cosmological observations, instead of the acceleration scale $a_0= c H_0$, which only de Sitter universe satisfies. 
Note that $a_0/6$ in Eq.~(\ref{a6}) with this $a_0$ deviates from Milgrom's constant by 30\%, which is bigger than the 10\% deviation of de Sitter $a_0$ value.
Thus, at first glance, the quasi de Sitter value may be worse than the de Sitter one.
However, we find later that this quasi de Sitter value of $a_0$ fits the galaxy rotation curves better than the $a_0$ value of the de Sitter case.

\section{Application to Galaxy Rotation Curves}\label{methods}

\subsection{Data}\label{data}

To examine the validity of Verlinde's emergent gravity, we compare the observed gravitational accelerations of galaxies with the predictions from the emergent gravity.
To do so, we have used the data in the {\it Spitzer} Photometry and Accurate Rotation Curves (SPARC) database, which contains a sample of 175 nearby disk galaxies with 3.6 $\mu$m surface photometry and rotation curves~\cite{Lelli:2016zqa}.
The SPARC team also provides mass models for all the galaxies along with the measured contribution of each component (i.e., gas disk, stellar disk, and bulge) to the observed rotation curve.
Among the 175 galaxies, we exclude 22 galaxies following the criteria of Ref.~\cite{Lelli:2016cui}:
10 face-on galaxies with small inclination ($i<30^\circ$) to minimize the effect of inclination correction on the rotational velocities, and 12 galaxies that have asymmetric rotation curves (quality flag $Q=3$) which may not trace the equilibrium gravitational potential.

\subsection{Results}\label{results}

For the remaining 153 galaxies, we first use their data to make a comparison between the observed gravitational acceleration derived from galactic rotation curves ($g_{\rm obs}=V_{\rm obs}^2/R$) and the Newtonian gravitational acceleration predicted from the baryon distribution ($g_{\rm bar}=V_{\rm bar}^2/R= (V_{\rm gas}^2 + \Upsilon_{\rm disk} V_{\rm disk}^2 + \Upsilon_{\rm bul} V_{\rm bul}^2)/R$) as in Refs.~\cite{McGaugh:2016leg, Lelli:2016cui}.
Here, $V_{\rm obs}$ is the rotational velocity of galaxy at galactocentric radius $R$, and $V_{\rm gas}$, $V_{\rm disk}$, and $V_{\rm bul}$ are the Newtonian gravitational contributions of gas disk, stellar disk, and bulge to the rotational velocity at the same galactocentric radius.
Following Ref.~\cite{Lelli:2016zqa}, we adopt the mass to light ratios for stellar disk and bulge as $\Upsilon_{\rm disk(bul)}=0.5(0.7)~ M_\odot/L_\odot$, respectively.

\begin{figure*}[t]
	\centering
	\includegraphics[width=0.95\linewidth]{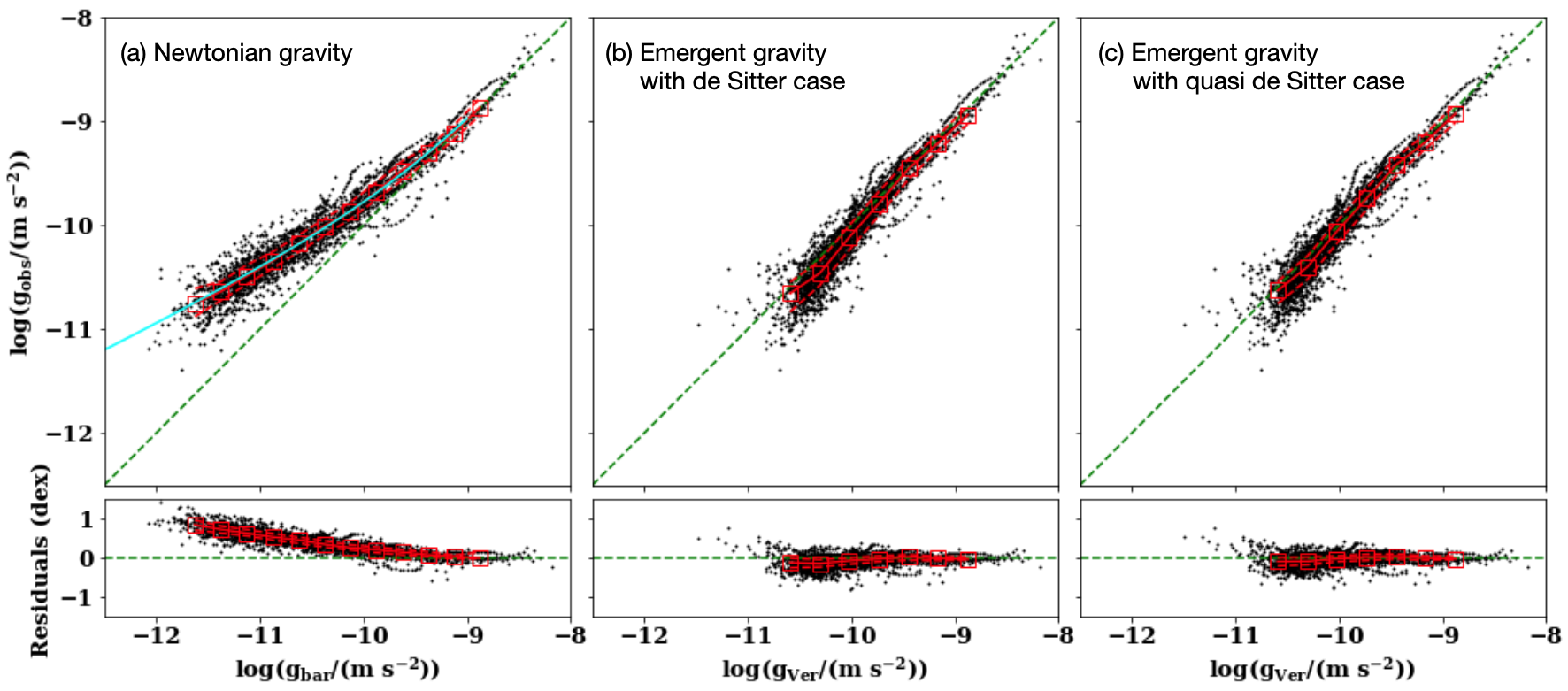}
	\caption{[Top ($a$)] The observed gravitational acceleration derived from galactic rotation curves ($g_{\rm obs}$) as a function of the Newtonian gravitational acceleration from baryon distribution ($g_{\rm bar}$).
		The red box indicates the median of the data (black dots) at each bin along with one sigma scatter denoted by upper and lower red dashed lines.
		The cyan solid and green dashed lines represent the best-fit~\cite{McGaugh:2016leg, Lelli:2016cui} and one-to-one relations, respectively.
		[Bottom ($a$)] The difference between the observed and the Newtonian gravitational accelerations as a function of $g_{\rm bar}$.
		[($b$)] Similar to ($a$), but as a function of the Verlinde gravitational acceleration ($g_{\rm Ver}$) with $a_0=c H_0$ (i.e., de Sitter case).
		[($c$)] Similar to ($b$), but with the quasi de Sitter value for $a_0$.
		\label{fig-gcomp}}
\end{figure*}

\begin{figure*}[t]
	\centering
	\includegraphics[width=0.9\linewidth]{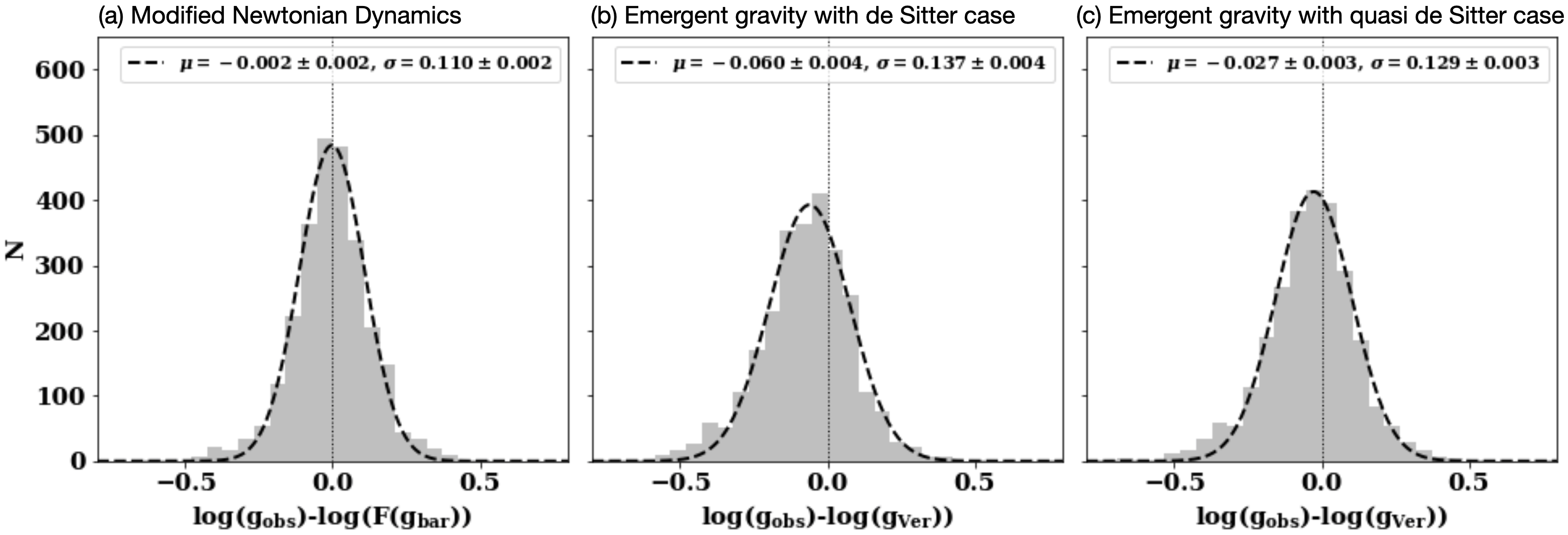}
	\caption{[($a$)] Histogram of the logarithmic difference between the observed and the predicted gravitational accelerations from the best-fit relation of Ref.~\cite{McGaugh:2016leg, Lelli:2016cui} as in the top of Fig.~\ref{fig-gcomp}($a$).
		The dashed line shows the Gaussian fit to the histogram along with the mean ($\mu$) and the standard deviation ($\sigma$).
		[($b$)] Similar to ($a$), but for the predicted acceleration from the emergent gravity as in the top of Fig.~\ref{fig-gcomp}(b) (i.e., de Sitter case with $a_0=c H_0$).
		[($c$)] Similar to ($b$), but with the quasi de Sitter value for $a_0$.
		\label{fig-ghist}}
\end{figure*}

Fig.~\ref{fig-gcomp}($a$) shows this comparison.
As already shown in Refs.~\cite{McGaugh:2016leg, Lelli:2016cui}, the observed and predicted gravitational accelerations are strongly correlated with a dispersion of $\sigma=0.11$ dex [see Fig.~\ref{fig-ghist}($a$)].
The offset between the two increases as the Newtonian gravitational acceleration from the baryon distribution decreases.
We also show the best-fit relation~\cite{McGaugh:2016leg, Lelli:2016cui} based on the fitting function suggested in Ref.~\cite{McGaugh:2008nc}:
\begin{equation}
g_{\mathrm{obs}}=\mathcal F(g_{\mathrm{bar}})=\frac{g_{\mathrm{bar}}}{1-e^{-\sqrt{{g_{\mathrm{bar}}}/g_\dagger}}}\,,
\label{McGaughMOND}
\end{equation}
where $g_\dagger$ is a free parameter with the best-fit value of $g_\dagger = 1.20\pm0.02$(random)$\pm~0.24$ (systematic)$\times10^{-10}~\mathrm{m~s^{-2}}$ for the SPARC sample.
The authors of Refs.~\cite{McGaugh:2016leg, Lelli:2016cui} considered this function because it smoothly connects the Newtonian acceleration and the MOND one in the two limits.
In other words, it satisfies the conditions (\ref{gobsconditions}).

In Fig.~\ref{fig-gcomp}($b$) and \ref{fig-gcomp}($c$), we show the observed gravitational acceleration as a function of the acceleration obtained from Verlinde's emergent gravity with Eq.~(\ref{Verlindegalaxy}).
Here, we need a density profile of baryonic matter $\rho_B$ and the radial gradients of baryonic gravitational acceleration $\partial_r g_B$.
The considered baryonic components in a galaxy are stellar and gas disks along with bulge.
For stellar and gas disks, we assume an exponential disk model of $\rho(R,z) = \rho_0 {\rm exp}(-R/R_d) {\rm exp}(-\left| z \right| / z_d)$ where $R_d$ and $z_d$ are scale length and height, respectively.\footnote{For more details, see Eq.~(1.10) of Ref.~\cite{Binney:2008bt}.}
We adopt the relevant parameters for stellar and gas disks from the surface photometry data of Ref.~\cite{Lelli:2016zqa}.
The scale height along the vertical distribution of the disk is assumed to be $z_d=0.196 R_d^{0.633}$~\cite{Bershady:2010sj}. Unlike MOND, which predicts convergent gravitational accelerations in the limit in which the width of galaxy goes to zero, despite the divergence of the mass density, Verlinde's emergent gravity predicts divergent gravitational accelerations in such a case. 
Therefore, it is necessary to introduce the width of galaxy in our analysis.
Fig.~\ref{fig-gcomp}($b$) and \ref{fig-gcomp}($c$) show that the observed and predicted gravitational accelerations are tightly correlated as in Fig.~\ref{fig-gcomp}($a$).
In addition, the data follow the one-to-one relation with a small offset.

To make a quantitative comparison between the observed and the predicted accelerations, we show histograms of the offsets for three predicted accelerations in Fig.~\ref{fig-ghist}: ($a$) from the fitting function $\mathcal F(g_{\mathrm{bar}})$~\cite{McGaugh:2008nc} and ($b$, $c$) from the emergent gravity.
The dashed line in each panel is the best-fit Gaussian curve, which was derived from the curve\textunderscore fit module in the PYTHON scipy.optimize package. The error for each parameter is the square root of the corresponding element on the diagonal component of the covariance matrix from the fit.
Fig.~\ref{fig-ghist}($a$) shows that the mean is consistent with zero with a scatter of $\sigma=0.11$, which agrees with the results of Refs.~\cite{McGaugh:2016leg, Lelli:2016cui}.
Fig.~\ref{fig-ghist}($b$) shows a slightly larger residual, but still small with $\mu=-0.060\pm0.004$ and $\sigma=0.137\pm0.004$.
Note that this is for the emergent gravity with $a_0= c H_0$ (de Sitter case).

As briefly explained at the end of Section~\ref{Verlindegravity}, the quasi de Sitter case appears closer to the real universe.
Thus, we also show the results for this case in Figs.~\ref{fig-gcomp}($c$) and ~\ref{fig-ghist}($c$).
When we compare Fig.~\ref{fig-ghist}($b$) and \ref{fig-ghist}($c$), the quasi de Sitter case shows a better agreement with observations as expected.
Not only the mean of the residuals in the quasi de Sitter is closer to zero than that in the de Sitter ($-0.027$ dex vs $-0.060$ dex), but also the standard deviation is smaller (0.129 dex vs 0.137 dex). 
The mean and the dispersion of the residuals even in the quasi de Sitter appear slightly larger than those for the fitting function of Ref.~\cite{McGaugh:2008nc}.
However, Eq.~(\ref{McGaughMOND}) is a simple empirical fitting function that does not have any physical background. 
Moreover, our value $a_0$, whether it is de Sitter value or quasi de Sitter one, is only an approximation, as we will explain in the next section. Thus, a correct value of $a_0$ has a possibility to yield different results.

\section{Discussion and Conclusions}\label{discussion}

We have shown that Verlinde's emergent gravity can explain the galaxy rotation curves well.
Considering the large errors from observations contributing to the residuals (0.12 dex, see Table 1 of Ref.~\cite{McGaugh:2016leg}), our small residual ($\mu=-0.027$ and $\sigma= 0.129$) for the quasi de Sitter case is remarkable as most of the residuals could be explained with the observational errors. Moreover, due to these large errors, it may be too premature to judge that the observational data prefers the MOND scenario to the Verlinde quasi de Sitter scenario with absolute certainty.

In any case, it is fair to say that the fitting function of Ref.~\cite{McGaugh:2008nc} based on MOND works equally well \cite{McGaugh:2016leg}, but the function does not have any physical background as mentioned earlier.
On the other hand, Verlinde's emergent gravity provides the physical and conceptual contexts of the formulae that agree with the observational data of galaxy rotation curves.
In the emergent gravity, the constant that plays the role of Milgrom's constant is $a_0/6$ which is independently determined from the observations that have nothing to do with the galaxy rotation curves.
The simple assumption of an additional volume contribution to the usual area law for the entanglement entropy explains the galaxy rotation curves.

There was a study suggesting that the emergent gravity does not agree with the rotation curves of dwarf galaxies~\cite{Pardo:2017jun}, which could be due to misinterpretation of $\Phi_B$ in the emergent gravity.
The $\Phi_B$ in Verlinde's work does not denote the Newtonian potential, but a quantity that appears in the space-space components of metric.
However, only the time-time component of metric is related to the Newtonian potential. 
Another study also claimed that the galaxy rotation curves do not match with the emergent gravity~\cite{Lelli:2017sul}.
However, they have used Verlinde's formula, which is correct only in the presence of spherical symmetry.
Certainly, the matter distributions in most galaxies lack spherical symmetry.
In this regard, our study using formulae that are also valid in the absence of spherical symmetry could be important.

There are also arguments that Verlinde's emergent gravity is not consistent with gravitational lensing data for galaxies or galaxy clusters~\cite{Luo:2020iup, ZuHone:2019hdt, Tamosiunas:2019ghq, Halenka:2018qnj}. 
However, note that the lensing data cannot be yet tested with the emergent gravity as its relativistic version is not still formulated.
Similarly, the emergent gravity still needs to pass other critical tests such as the CMB (cosmic microwave background) anisotropy, which is possible only when the relativistic version of the emergent gravity is formulated.

We also would like to remark that $a_0$ in the quasi de Sitter model is based on the $\Lambda$CDM model.
Of course, if Verlinde's emergent gravity is correct, we may not need cold dark matter.
However, at present, there is no other way to obtain $a_0$, unless we assume that our universe is de Sitter, which is certainly not true; we need some kind of model regarding the accelerated expansion of our universe.
$\Lambda$CDM is so far the best model regarding it.
This is another reason why we need the relativistic formulation of the emergent gravity.
The quasi de Sitter model considered in Ref.~\cite{Diez-Tejedor:2016fdn} to calculate $a_0$ is yet incomplete.
If the final version of the emergent gravity is formulated, we will be able to obtain the real $a_0$.
The quasi de Sitter value for $a_0$ is, at best, an approximation.

Aside from this determination of $a_0$ based on the relativistic formulation of the emergent gravity and cosmological observations, the measurement of the violation of equivalence principle will be helpful for determining $a_0$.
In Ref.~\cite{Yoon:2021bse}, it was shown that Verlinde's emergent gravity predicts the violation of equivalence principle, and this violation is proportional to $a_0$.
As explained in Ref.~\cite{Yoon:2021bse}, the gravitational acceleration depends on the density of test particle, according to Verlinde's emergent gravity.
However, experiments so far could not detect the violation of equivalence principle, because they mainly concerned with the other criteria that may cause the difference in the gravitational acceleration. 
For example, the authors of Ref.~\cite{Schlamminger:2007ht} measured the gravitational acceleration difference between beryllium and titanium, because their baryon number difference is big.
However, Ref.~\cite{Yoon:2021bse} indicates that the gravitational acceleration difference between beryllium and gold is more than 60 times bigger than the gravitational acceleration between beryllium and titanium.
Thus, the gravitational acceleration difference is big enough to be noticeable if the experiment in Ref.~\cite{Schlamminger:2007ht} is performed again with beryllium and gold instead, at the same experimental sensitivity.

Currently, there is a broad consensus that many phenomena cannot be explained without assuming the existence of dark matter.
However, most, if not all, of them are based on the assumption that Newton-Einstein theory of gravity is correct.
Therefore, it is important to look for a new theory of gravity that is also natural and conceptually simple from physical point of view.
Of course, Verlinde's emergent gravity does not explain many phenomena yet, except galaxy rotation curves, so further research is necessary.

\ack

The work is supported by the National Research Foundation of Korea (NRF) [NRF-2019R1C1C1005073 and NRF-2021R1A4A2001897 (YY, JCP), NRF-2021R1A2C1094577 (HSH)].

\section*{References}

\bibliographystyle{iopart-num}
\bibliography{ref}

\providecommand{\newblock}{}
\begin{thebibliography}{10}
\expandafter\ifx\csname url\endcsname\relax
  \def\url#1{{\tt #1}}\fi
\expandafter\ifx\csname urlprefix\endcsname\relax\def\urlprefix{URL }\fi
\providecommand{\eprint}[2][]{\url{#2}}

\bibitem{Zwicky:1933gu}
Zwicky F 1933 {\em Helv. Phys. Acta\/} {\bf 6} 110--127

\bibitem{Trippe:2014hja}
Trippe S 2014 {\em Z. Naturforsch. A\/} {\bf 69} 173 (\textit{Preprint}
  \eprint{1401.5904})

\bibitem{Bullock:2017xww}
Bullock J~S and Boylan-Kolchin M 2017 {\em Ann. Rev. Astron. Astrophys.\/} {\bf
  55} 343--387 (\textit{Preprint} \eprint{1707.04256})

\bibitem{DiValentino:2021izs}
Di~Valentino E, Mena O, Pan S, Visinelli L, Yang W, Melchiorri A, Mota D~F,
  Riess A~G and Silk J 2021 {\em Class. Quant. Grav.\/} {\bf 38} 153001
  (\textit{Preprint} \eprint{2103.01183})

\bibitem{Tully:1977fu}
Tully R~B and Fisher J~R 1977 {\em Astron. Astrophys.\/} {\bf 54} 661--673

\bibitem{Milgrom:1983ca}
Milgrom M 1983 {\em Astrophys. J.\/} {\bf 270} 365--370

\bibitem{Lelli:2022hvc}
Lelli F 2022 {\em Nature Astron.\/} {\bf 6} 35--47 (\textit{Preprint}
  \eprint{2201.11752})

\bibitem{McGaugh:2016leg}
McGaugh S, Lelli F and Schombert J 2016 {\em Phys. Rev. Lett.\/} {\bf 117}
  201101 (\textit{Preprint} \eprint{1609.05917})

\bibitem{Lelli:2016cui}
Lelli F, McGaugh S~S, Schombert J~M and Pawlowski M~S 2017 {\em Astrophys.
  J.\/} {\bf 836} 152 (\textit{Preprint} \eprint{1610.08981})

\bibitem{Verlinde:2016toy}
Verlinde E~P 2017 {\em SciPost Phys.\/} {\bf 2} 016 (\textit{Preprint}
  \eprint{1611.02269})

\bibitem{Yoon:2021bse}
Yoon Y 2021 {\em Int. J. Mod. Phys. D\/} {\bf 30} 2150024

\bibitem{Lelli:2015wst}
Lelli F, McGaugh S~S and Schombert J~M 2016 {\em Astrophys. J. Lett.\/} {\bf
  816} L14 (\textit{Preprint} \eprint{1512.04543})

\bibitem{McGaugh:2000sr}
McGaugh S~S, Schombert J~M, Bothun G~D and de~Blok W~J~G 2000 {\em Astrophys.
  J. Lett.\/} {\bf 533} L99--L102 (\textit{Preprint} \eprint{astro-ph/0003001})

\bibitem{Dalcanton:1996de}
Dalcanton J~J, Spergel D~N and Summers F~J 1997 {\em Astrophys. J.\/} {\bf 482}
  659--676 (\textit{Preprint} \eprint{astro-ph/9611226})

\bibitem{Mo:1997vb}
Mo H~J, Mao S and White S~D~M 1998 {\em Mon. Not. Roy. Astron. Soc.\/} {\bf
  295} 319 (\textit{Preprint} \eprint{astro-ph/9707093})

\bibitem{Mo:2000pu}
Mo H~J and Mao S 2000 {\em Mon. Not. Roy. Astron. Soc.\/} {\bf 318} 163
  (\textit{Preprint} \eprint{astro-ph/0002451})

\bibitem{Begum:2008gn}
Begum A, Chengalur J~N, Karachentsev I~D and Sharina M~E 2008 {\em Mon. Not.
  Roy. Astron. Soc.\/} {\bf 386} 138 (\textit{Preprint} \eprint{0801.3606})

\bibitem{Kirby:2012sg}
Kirby E~M, Koribalski B, Jerjen H and Lopez-Sanchez A 2012 {\em Mon. Not. Roy.
  Astron. Soc.\/} {\bf 420} 2924 (\textit{Preprint} \eprint{1202.0354})

\bibitem{McCall:2012ya}
McCall M~L, Vaduvescu O, Nunez F~P, Dominguez A~B, Fingerhut R, Unda-Sanzana E,
  Li B and Albrecht M 2012 {\em Astron. Astrophys.\/} {\bf 540} A49
  (\textit{Preprint} \eprint{1204.1074})

\bibitem{Brook:2016}
Brook C~B, Santos-Santos I and Stinson G 2016 {\em Monthly Notices of the Royal
  Astronomical Society\/} {\bf 459} 638 (\textit{Preprint} \eprint{1603.06595})

\bibitem{Karachentsev:2016cgo}
Karachentsev I~D, Kaisina E~I and Kashibadze O~G 2016 {\em Astron. J.\/} {\bf
  153} 6 (\textit{Preprint} \eprint{1611.02574})

\bibitem{Seeram:2021xsi}
Seeram S and Desai S 2021 {\em J. Astrophys. Astron.\/} {\bf 42} 3

\bibitem{Clifton:2011jh}
Clifton T, Ferreira P~G, Padilla A and Skordis C 2012 {\em Phys. Rept.\/} {\bf
  513} 1--189 (\textit{Preprint} \eprint{1106.2476})

\bibitem{Verlinde:2010hp}
Verlinde E~P 2011 {\em JHEP\/} {\bf 04} 029 (\textit{Preprint}
  \eprint{1001.0785})

\bibitem{Yoon:2020meh}
Yoon Y 2020 {Comment on ''Inconsistencies in Verlinde's emergent gravity''}
  (\textit{Preprint} \eprint{2003.03198})

\bibitem{Diez-Tejedor:2016fdn}
Diez-Tejedor A, Gonzalez-Morales A~X and Niz G 2018 {\em Mon. Not. Roy. Astron.
  Soc.\/} {\bf 477} 1285--1295 (\textit{Preprint} \eprint{1612.06282})

\bibitem{Lelli:2016zqa}
Lelli F, McGaugh S~S and Schombert J~M 2016 {\em Astron. J.\/} {\bf 152} 157
  (\textit{Preprint} \eprint{1606.09251})

\bibitem{McGaugh:2008nc}
McGaugh S 2008 {\em Astrophys. J.\/} {\bf 683} 137--148 (\textit{Preprint}
  \eprint{0804.1314})

\bibitem{Binney:2008bt}
{Binney} J and {Tremaine} S 2008 {\em {Galactic Dynamics: Second Edition}\/}

\bibitem{Bershady:2010sj}
{Bershady} M~A, {Verheijen} M~A~W, {Westfall} K~B, {Andersen} D~R, {Swaters}
  R~A and {Martinsson} T 2010 {\em Astrophys. J.\/} {\bf 716} 234--268
  (\textit{Preprint} \eprint{1004.5043})

\bibitem{Pardo:2017jun}
Pardo K 2020 {\em JCAP\/} {\bf 12} 012 (\textit{Preprint} \eprint{1706.00785})

\bibitem{Lelli:2017sul}
Lelli F, McGaugh S~S and Schombert J~M 2017 {\em Mon. Not. Roy. Astron. Soc.\/}
  {\bf 468} L68--L71 (\textit{Preprint} \eprint{1702.04355})

\bibitem{Luo:2020iup}
Luo W, Zhang J, Halenka V, Yang X, More S, Miller C, Sunayama T, Liu L and Shi
  F 2021 {\em Astrophys. J.\/} {\bf 914} 96 (\textit{Preprint}
  \eprint{2003.09818})

\bibitem{ZuHone:2019hdt}
ZuHone J~A and Sims J~R 2019 {\em Astrophys. J.\/} {\bf 880} 145
  (\textit{Preprint} \eprint{1905.03832})

\bibitem{Tamosiunas:2019ghq}
Tamosiunas A, Bacon D, Koyama K and Nichol R~C 2019 {\em JCAP\/} {\bf 05} 053
  (\textit{Preprint} \eprint{1901.05505})

\bibitem{Halenka:2018qnj}
Halenka V and Miller C~J 2020 {\em Phys. Rev. D\/} {\bf 102} 084007
  (\textit{Preprint} \eprint{1807.01689})

\bibitem{Schlamminger:2007ht}
Schlamminger S, Choi K~Y, Wagner T~A, Gundlach J~H and Adelberger E~G 2008 {\em
  Phys. Rev. Lett.\/} {\bf 100} 041101 (\textit{Preprint} \eprint{0712.0607})

\end{thebibliography}

\end{document}